\documentclass[%
 reprint,
superscriptaddress,
 amsmath,amssymb,
 aps,
 onecolumn,
 pre,
floatfix,
]{revtex4-2}

\usepackage{graphicx}
\usepackage{dcolumn}
\usepackage{bm}
\usepackage{xcolor}
\usepackage[caption=false]{subfig}
\usepackage{hyperref}

\usepackage{cleveref}
\crefformat{section}{\S#2#1#3}

\begin{document}

\preprint{APS/123-QED}

\title{Biolocomotion and premelting in ice}

\author{J\' er\' emy Vachier}
\email{jeremy.vachier@su.se}
\affiliation{Nordita, KTH Royal Institute of Technology and Stockholm University, Hannes Alfv\' ens v\" ag 12, SE-106 91 Stockholm, Sweden}
\author{J. S. Wettlaufer}
\email{john.wettlaufer@yale.edu}
\affiliation{Nordita, KTH Royal Institute of Technology and Stockholm University, Hannes Alfv\' ens v\" ag 12, SE-106 91 Stockholm, Sweden}
\affiliation{%
 Yale University, New Haven, Connecticut 06520-8109, USA \\
 }%

\date{\today}

\begin{abstract}
Biota are found in glaciers, ice sheets and permafrost. Ice bound micro-organisms evolve in a complex mobile environment facilitated or hindered by a range of bulk and surface interactions. When a particle is embedded in a host solid near its bulk melting temperature, a melted film forms at the surface of the particle in a process known as interfacial premelting. Under a temperature gradient, the particle is driven by a thermomolecular pressure gradient toward regions of higher temperatures in a process called thermal regelation. When the host solid is ice and the particles are biota, thriving in their environment requires the development of strategies, such as producing exopolymeric substances (EPS) and antifreeze glycoproteins (AFP) that enhance the interfacial water. Therefore, thermal regelation is enhanced and modified by a process we term {\em bio-enhanced premelting}. Additionally, the motion of bioparticles is influenced by chemical gradients influenced by nutrients within the icy host body. We show how the overall trajectory of bioparticles is controlled by a competition between thermal regelation and directed biolocomotion.  By re-casting this class of regelation phenomena in the stochastic framework of active Ornstein-Uhlenbeck dynamics, and using multiple scales analysis, we find that for an attractive (repulsive) nutrient source, that thermal regelation is enhanced (suppressed) by biolocomotion. This phenomena is important in astrobiology, the biosignatures of extremophiles and in terrestrial paleoclimatology.  
\end{abstract}

\maketitle

\section{Introduction}
\indent Ice sheets are an essential reservoir of information on past climate and they contain an important record of micro-organisms on Earth, recording ice microbes and their viruses over long periods \cite{karl-science-1999,christner-enm-2001}. In these extreme environments, the abundance of virus is well correlated with bacterial abundance, but is 10 to 100 times lower than in temperate aquatic ecosystems \cite{anesio-jgrb-2007}. Even in these harsh conditions, the virus infection rate is relatively high \cite{bellas-fm-2015}, leading to the expectation of low long-term survival rates.
However, recent studies have shown that some viruses develop survival strategies to maintain a long-term relationship with their hosts \cite{bellas-fm-2015,heilmann-pnas-2012}, possibly up to thousands of years \cite{zhong-m-2021}. For example, viruses such as bacteriophages can switch to a lysogenic life strategy enabling them to replicate and maintain themselves in the bacterial population without lysis over multiple generations \cite{bellas-fm-2015}. Moreover, among these viruses some can provide immunity to their hosts against other viruses \cite{bellas-fm-2015,yau-pnas-2011}, or manipulate their metabolism to facilitate nutrient acquisition by affecting motility genes \cite{zhong-m-2021}.  Indeed, motile biota are found to be active in ice for substantial periods. For example, recently a 30,000 year old giant virus {\textit{Pithovirus sibericum}} was found in permafrost \cite{legendre-pnas-2014} along with microbes \cite{zhong-micro-2021,el-espr-2021} and nematodes \cite{shatilovich-book-2018}, and viable bacteria have been found in 750,000 year old glacial ice \cite{christner-em-2003}. 
Basal ice often contains subglacial debris and sediment, which serve as a source of nutrients and organic matter, providing a habitat for micro-organisms adapted to subfreezing conditions \cite{doyle-bioarxiv-2021,anesio-nature-2017}. Additionally, the microbiomes of sediment rich basal ices are distinct from those found in glacial ice and are equivalent to those found in permafrost \cite{doyle-bioarxiv-2021}, expanding the nature of subfreezing habitats.  
\newline
\indent Ice cores provide the highest resolution records of past climate states \cite[e.g.,][]{royer-springer-1983, legrand-rg-1997, stauffer-ag-2004, Alley:2010, thomas-geobiol-2015,tetzner-fes-2021}.  Of particular relevance to our study is their role as a refuge for micro-organisms, from the recent past \cite[][]{papina-cp-2013, mao-fm-2022} to millennia \cite{achberger-fm-2011,knowlton-bio-2013,garcia-fm-2021}.
Ice microbes are taxonomically diverse and have a wide range of taxonomic relatives \cite{anesio-nature-2017,wilhelm-isme-2013,garcia-mms-2021,stibal-emr-2015,knowlton-bio-2013}. Common algae taxa are centric and pennate diatoms, dinoflagellates and flagellates \cite{hop-mms-2020,kauko-mms-2018,spilling-mms-2018}, whereas common bacterial taxa are pseudomonadota, actinobacteria, firmicutes and bacteroidetes \cite{zhong-m-2021,itcus-sr-2018}. Many of these microbes have different motility mechanisms \cite{miyata-gc-2020,hahnke-fm-2016} from swimming (e.g., \textit{Chlamydomonas
nivalis} \cite{hill-jtb-1997} or \textit{Methylobacterium} \cite{tsagkari-water-2018,doerges-Naturwissenschaften-2014,zhong-m-2021}) to gliding (e.g., \textit{diatoms} \cite{svensson-plosone-2014,aumack-jms-2014} \textit{or Bacillus subtilis} \cite{knowlton-bio-2013,christner-icarus-2000}), which can be used to assess their locomotion. 
Examples of biological proxies include diatoms \cite{biswas-jb-2021} and bacteria colonies \cite{dong-gsa-2010,delgado-nbe-2017}, reflecting a unique range of physical-biological interactions in the climate system. Therefore, understanding the relationship between the evolution of ice bound micro-organisms and  proxy dating methods is a key aspect of understanding the covariation of life and climate. 
\newline
Finally, such understanding is essential for the study of extraterrestrial life. In our own solar system, despite the debate regarding the existence of bulk water on Mars \cite[e.g.,][]{Benningfield:2022}, thin water films, such as those studied here, hold the most potential for harboring life under extreme conditions. Indeed, lipids, nucleic acids, and amino acids influenced by active motility may serve as biosignatures of extra terrestial life.  Combining measurements of diffusivity-characterized-motility \cite{lindensmith-po-2016,nadeau-astro-2016} with bioparticle distribution observed on Earth, provides crucial information for development of new instrumentation to detect the presence of extra terrestrial life \cite{nadeau-astro-2016,jones-frontiersmico-2018}. Indeed, recently micro-organisms trapped in primary fluid inclusions in halite for millions of years have been discovered \cite{schreder-geology-2021}, providing promise for both terrestrial and extraterrestrial biosignature detection.
\begin{figure}
    \centering
    \includegraphics[scale=.6]{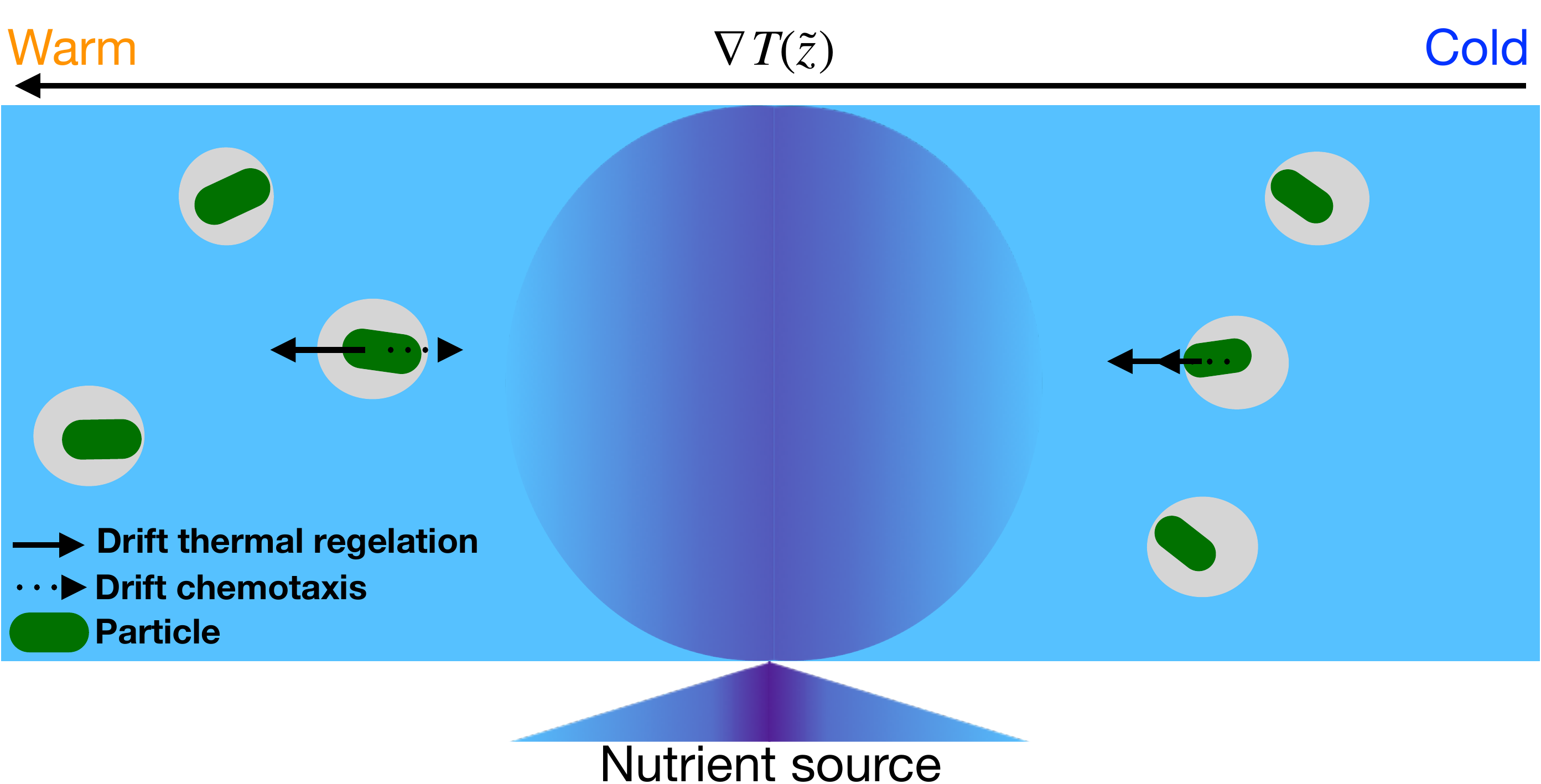}
    \caption{Schematic of few active particles embedded in ice under an external temperature gradient $\nabla T$ along the $\tilde{z}$-axis. The nutrient source is shown by the purple gradient. The external temperature gradient induces a drift velocity, and particles move toward regions of higher temperature, in a process known as thermal regelation (black arrow).  An additional drift velocity is associated with particle motion towards higher concentration of nutrients (black dotted arrow).  Thus, depending on a particle's position, and the background temperature and nutrient gradients, these two drift effects can compete or amplify each other.}
    \label{fig1:intro}
\end{figure}
\newline
\indent When a particle is embedded in ice near the bulk melting temperature, the ice may melt at the particle-ice surface in a process known as \textit{interfacial premelting} \cite{dash-rmp-2006physics}. The thickness of the melt film depends on the temperature, impurities, material properties and geometry. A temperature gradient is accompanied by a thermomolecular pressure gradient that drives the interfacial liquid from high to low temperatures, and hence the particle migrates from low to high temperatures in a process called \textit{thermal regelation} \cite{rempel-prl-2001, dash-rmp-2006physics,wettlaufer-arfm-2006,peppin-jsp-2009,marath-sm-2020}.
Thermal regelation of inert particles plays a major role in the redistribution of material inside of ice, which has important environmental and composite materials implications \cite{rempel-prl-2001, dash-rmp-2006physics,wettlaufer-arfm-2006,peppin-jsp-2009,marath-sm-2020}. Moreover, surface properties are central to the fact that extremophile organisms in Earth's cryosphere--glaciers, sea ice and permafrost--develop strategies to persist in challenging environments.   Indeed, many biological organisms secrete exopolymeric substance (EPS) \cite{wingender-book-1999} or harness antifreeze glycoproteins (AFP) \cite{bang-md-2013,eskandari-bio-2020} to maintain interfacial liquidity.  For example, sea ice houses an array of algae and bacteria, some of which produce EPS to protect them at low temperature and high salinity \cite{ewert-fme-2014,ewert-biol-2013}.  Additionally, the enhanced liquidity associated with high concentrations of EPS alters the physical properties of sea ice and thereby play a role in climate change \cite{krembs-pnas-2011,decho-fm-2017}.
\newline
\indent In bulk solution, active particles act as simple microscopic models for living systems and are particularly accurate at mimicking the propulsion of bacteria or algae \citep[e.g.,][]{cates-rpp-2012,bechinger-rmp-2016,ghosh-nl-2009,kim-am-2013,jin-pre-2019}. By converting energy to motion using biological, chemical, or physical processes, they exhibit rich collective emergent motion from ostensibly simple rules 
\citep[e.g.,][]{elgeti-rpp-2015,romanczuk-epjst-2012}.   Algae and bacteria operate in complex geometries and translate environmental conditions into microscopic information that guides their behavior. 
 Examples of such information include quorum sensing (e.g., particle population density), used by bacteria to regulate biofilm formation, defense against competitors and adapt to changing environments \cite{li-sensors-2012,yan-fp-2019,lee-mbio-2020}; chemotaxis (e.g., concentration gradients of nutrients), used by algae/bacteria to direct their motion toward higher concentrations of beneficial, or lower concentrations of toxic, chemicals \cite{wadhams-nature-2004,showalter-envi-2018,cremer-nature-2019,bar-jrsi-2016,mattingly-natphy-2021}.  It is important to emphasize that factors such as surface adhesion, salinity, the segregation of impurities of all types from the ice lattice, among other factors \cite[See e.g.,][]{Bar-Dolev:2012, hansen-pss-2014, bar-jrsi-2016, showalter-emr-2018}, make our treatment of chemotaxis a simplified starting point.   
However, field samples and laboratory experiments have shown that cell motility is influenced by chemotaxis at low temperature \cite{junge-aem-2003,lindensmith-po-2016, mudge-em-2021}.  Thus, although there are many complicated interactions that provide scope for future work, the basic role of chemotaxis in the distribution of biota in ice must start with a self-consistent framework, which is the focus of our work.
\newline
\indent The confluence of thermal regelation, bio-enhanced premelting and intrinsic mobility underlie our study. Indeed, intrinsic mobility and chemotaxis  may compete with thermal regelation, which constitutes a new area of research--ice bound active particles in premelting ice, as illustrated in Fig. \ref{fig1:intro}.  Moreover, including micro-organism protection mechanisms that enhance interfacial liquidity, such as the secretion of EPS, constitute a unique class of regelation phenomena. Finally, treating this corpus of processes quantitatively is particularly relevant for climatology and the global carbon cycle \citep[e.g.,][]{wadham-naturecom-2019,holland-bio-2019}.
\newline
\indent Our framework is the active Ornstein-Uhlenbeck particle (AOUP) \citep[e.g][]{fodor-prl-2016,caprini-sm-2018,martin-pre-2021,dabelow-fp-2021,caprini-sr-2019,bonilla-pre-2019,dabelow-prx-2019}. The active force is governed by an Ornstein-Uhlenbeck process with magnitude $\tilde{D}_a$, which is the active diffusivity. This force can be compared to a colored noise process \cite{martin-pre-2021,sevilla-pre-2019}. In addition to the active diffusivity, the AOUP is characterized by a time $\tau_a$, which defines the noise persistence, from which the system switches from a ballistic to a diffusive regime. The active diffusivity $\tilde{D}_a$ and characteristic time $\tau_a$ can be measured experimentally \cite{maggi-prl-2014,maggi-sr-2017,donado-sr-2017}. 
The AOUP has been shown to provide accurate predictions for a range of complex phenomena \cite{marconi-sm-2016,caprini-sm-2018,caprini-sr-2019,dabelow-fp-2021}, and is theoretically advantageous due to its Gaussian nature \cite{martin-pre-2021}. 
These issues motivate our use of the AOUP model framework to describe the motion of active particles in ice under an external temperature gradient with a nutrient source. We analyze these particles in three dimensions using a multiple scale expansion to derive the associated Fokker-Planck equation. \newline
\indent The paper is organized as follows.  In order to make our treatment reasonably self-contained we note that we are generalizing our previous approach \cite{marath-sm-2020,vachier-pre-2022}, which we recover in the appropriate limit.  Thus, in \S \ref{sec:method} we introduce the active Ornstein-Uhlenbeck model for bio-premelted particles and in \S \ref{sec:results} we derive the associated Fokker-Planck equation using a multiple scale expansion.  We then compare our analytic and numerical solutions after which, in \S \ref{sec:conclusion}, we draw conclusions.  

\section{Methods}\label{sec:method}
Thermal regelation is  understood as a consequence of the premelted film around a particle, originally treated as {\em inert}, that (a) executes diffusive motion in the ice column with diffusivity $\tilde{D}(\tilde{z})\mathbb{I}$, where $\mathbb{I}$ is the identity matrix, and (b) experiences a drift velocity $\tilde{v}(\tilde{z})=U(\tilde{z})\hat{\tilde{\bm{z}}}$ parallel to the temperature gradient \cite{marath-sm-2020}.
Therefore, regelation biases the motion of an {\em active particle} by the drift velocity $U(\tilde{z})\hat{\tilde{\bm{z}}}$.
\newline
\indent For inert particles with a sufficiently large number of moles of electrolyte impurities per unit area of surface, $N_i$, the premelted film thickness $d \propto N_i$ \cite{wettlaufer-prl-1999, marath-sm-2020}.
However, the production of EPS/AFP enhances liquidity at the ice surface by increasing the impurity concentration \cite{krembs-pnas-2011, ewert-bio-2013,hansen-pss-2014,anesio-nature-2017}, which 
we treat here using an enhancement factor as $N=nN_i$, which gives 
 \begin{equation}
d = \frac{R_g T_m^2 N}{\rho_l q_m\Delta T}\,,
\label{eq:thickness2}
\end{equation}
where the universal gas constant is $R_g$,  the latent heat of fusion per mole of the solid is $q_m$,  the molar density  of the liquid is $\rho_l$, the undercooling is $\Delta T = T_m-T$ with $T_m= 273.15$K the pure bulk melting temperature and $T$ the temperature of ice. 
\newline
\indent The velocity and premelting-controlled
diffusivity are given by 
\begin{equation}
U(\tilde{z}) = - \frac{A_3}{A_2^3}\frac{1}{\tilde{z}^3}\, \qquad \text {and}
\label{eq:drift}
\end{equation}
\begin{equation}
\tilde{D}(\tilde{z}) = \frac{(R_gT_mN)^3}{8\pi \nu R^4 A_2^3}\frac{k_BT_m}{\tilde{z}^3}\,,
\label{eq:diff}
\end{equation}
respectively, where $A_2= \rho_lq_m\frac{|\nabla T|}{T_m}$ and $A_3 = \rho_sq_m|\nabla T|\frac{(R_gT_mN)^3}{6\nu RT_m}$, with $|\nabla T|$ the external temperature gradient. The viscosity of the fluid is $\nu$, the particle radius is $R$ and $k_B$ is the Boltzmann constant. Here, $\rho_sq_m\sim 334 \times 10^{6}$ J m$^{-3}$ \cite{marath-sm-2020}.
The evolution of the particle position $\tilde{\bm{r}}=(\tilde{r}_1,\tilde{r}_2,\tilde{r}_3)=(\tilde{x},\tilde{y},\tilde{z})$ and its activity are described by two overdamped Langevin equations
\begin{align}
\label{eq:lange_posi}
\frac{d}{d\tilde{t}} \tilde{\bm{r}}(\tilde{t}) &= \beta_D\nabla_{\tilde{\bm{r}}}\tilde{\mathcal{C}}(\tilde{\bm{r}},\tilde{t})+\sqrt{2\tilde{D}_a}\tilde{\bm{\eta}}+\tilde{v}(\tilde{z})+\sqrt{2\tilde{D}(\tilde{z})}\bm{\xi}_p(\tilde{t})\qquad \text{and}\\
\label{eq:lange_eta}
\frac{d}{d\tilde{t}}\tilde{\bm{\eta}}(\tilde{t}) &= -\frac{1}{\tau_a}\tilde{\bm{\eta}}(\tilde{t}) + \frac{1}{\tau_a}\bm{\xi}_a(\tilde{t}) \,.
\end{align}
The first term on the right-hand side of Eq. \eqref{eq:lange_posi} treats the chemotaxis response, representing the effect of the nutrient source of concentration $\tilde{\mathcal{C}}$ on the particle dynamics, 
where $\beta_D$ is the chemotactic strength \cite{keller-jtb-1971, saha-pre-2014,liebchen-acr-2018,pohl-prl-2014}, which we treat as a constant determined by the parameters in our system.  We note, however, that the transport properties of sea- and glacial-ice depend on their unique phase fraction  evolution \cite[e.g.][]{WWH97, rempel-nature-2001, Rempel-JGR-2002, WWO-2010}, which would clearly influence the effective--porosity dependent $\beta_D$.
In the ideal case, wherein the nutrient source is isotropic and purely diffusive, we have
\begin{equation}
\frac{\partial}{\partial \tilde{t}} \tilde{\mathcal{C}}(\tilde{\bm{r}},\tilde{t}) = \tilde{D}_{ch}\nabla^2_{\tilde{\bm{r}}}\tilde{\mathcal{C}}(\tilde{\bm{r}},\tilde{t})\,,
\label{eq:nutrients_field}
\end{equation}
where  $\tilde{D}_{ch}$ is the nutrient diffusivity. The activity, or self-propulsion, is given by the term $\sqrt{2\tilde{D}_a}\tilde{\bm{\eta}}$ in Eq. \eqref{eq:lange_posi}, with $\tilde{D}_a$ the active diffusivity.   The latter represents the active fluctuations of the system, such as those originating in particular processes described in Refs. \cite[][]{joanny-prl-2003,peruani-prl-2007, romanczuk-prl-2011,vandebroek-sm-2017}. 
Nutrient sources, such as dissolved silica, oxygen, nitrogen and methane, play a vital role in the life of ice-bound micro-organisms, such as algae and bacteria \cite[e.g.,][]{price-pnas-2000,Campen:2003,tung-astro-2006,mader-geo-2006, Rohde-pnas-2007,vancoppenolle-geo-2010}. Here we assume that $\tilde{D}_{ch}>\tilde{D}_a$, consistent with \cite{wu-aem-2006,amar-sc-2016,waahlin-crst-2017}, and $\tilde{D}_{ch}>\tilde{D}(\tilde{z})$. The function $\tilde{\bm{\eta}}=(\tilde{\eta}_1,\tilde{\eta}_2,\tilde{\eta}_3)$ is described by an Ornstein-Uhlenbeck process, with correlations given by
\begin{equation}
\langle\tilde{\eta_i}(\tilde{t}')\tilde{\eta_j}(\tilde{t}) \rangle = \frac{\delta_{ij}}{{2}\tau_a}e^{-\frac{|\tilde{t}'-\tilde{t}|}{\tau_a}}\,,
\label{eq:OU}
\end{equation}
where $\tau_a$ is the noise persistence as noted above. In the small $\tau_a$ limit, $\tilde{\bm{\eta}}$ reduces to  Gaussian white noise with correlations $\langle \tilde{\eta}_i(t')\tilde{\eta}_j(t)\rangle = \delta_{ij}\delta(\tilde{t}'-\tilde{t})$. In contrast, $\tilde{\bm{\eta}}$ does not reduce to Gaussian white noise when $\tau_a$ is finite, and Eq. \eqref{eq:lange_posi} does not reach equilibrium. Hence, $\tau_a$ controls the non-equilibrium properties of the dynamics \cite{martin-pre-2021,dabelow-fp-2021}.  
Finally, the random fluctuations in Eqs. \eqref{eq:lange_posi} and \eqref{eq:lange_eta}
are given by zero mean Gaussian white noise processes $\langle \xi_{p_i}(\tilde{t}')\xi_{p_j}(\tilde{t}) \rangle = \delta_{ij}\delta(\tilde{t}'-\tilde{t})$ and $\langle \xi_{a_i}(\tilde{t}')\xi_{a_j}(\tilde{t}) \rangle = \delta_{ij}\delta(\tilde{t}'-\tilde{t})$. \newline
\indent The Langevin Eqs. \eqref{eq:lange_posi} and \eqref{eq:lange_eta}, allow us to express the probability of finding a particle at the position $\tilde{\bm{r}}=(\tilde{r}_1,\tilde{r}_2,\tilde{r}_3)=(\tilde{x},\tilde{y},\tilde{z})$ at a given time $\tilde{t}$ through the Fokker-Planck equation, which describes the evolution of
the probability density function $P(\tilde{\bm{r}},\tilde{\bm{\eta}},\tilde{t}|\tilde{\bm{r}}_0,\tilde{\bm{\eta}}_0,\tilde{t}_0)$, with the initial condition $P(\tilde{\bm{r}},\tilde{\bm{\eta}},\tilde{t}=\tilde{t}_0|\tilde{\bm{r}}_0,\tilde{\bm{\eta}}_0,\tilde{t}_0)=\delta(\tilde{\bm{r}}-\tilde{\bm{r}}_0)\delta(\tilde{\bm{\eta}}-\tilde{\bm{\eta}}_0)$. To simplify the notation, we write the conditional probability as $P(\tilde{\bm{r}},\tilde{\bm{\eta}},\tilde{t}|\tilde{\bm{r}}_0,\tilde{\bm{\eta}}_0,\tilde{t}_0) \equiv P(\tilde{\bm{r}},\tilde{\bm{\eta}},\tilde{t})$ and eventually arrive at the following system of coupled
equations, 
{\small
\begin{align}
\label{eq:fpe3D}
\frac{\partial}{\partial \tilde{t}}  P(\tilde{\bm{r}},\tilde{\bm{\eta}},\tilde{t})&= -\beta_D\nabla_{\tilde{\bm{r}}} {\cdot} \left[ P(\tilde{\bm{r}},\tilde{\bm{\eta}},\tilde{t}) \nabla_{\tilde{\bm{r}}}\tilde{\mathcal{C}}(\tilde{\bm{r}},\tilde{t})\right] -\frac{\partial}{\partial \tilde{r}_3}\left[ \tilde{v}(\tilde{r}_3) P(\tilde{\bm{r}},\tilde{\bm{\eta}},\tilde{t})\right]-\sqrt{2\tilde{D}_a}\tilde{\bm{\eta}}\cdot\nabla_{\tilde{\bm{r}}} P(\tilde{\bm{r}},\tilde{\bm{\eta}},\tilde{t})\nonumber\\
&+\nabla^2_{\tilde{\bm{r}}}\left[\tilde{D}(\tilde{r}_3) P(\tilde{\bm{r}},\tilde{\bm{\eta}},\tilde{t})\right]+\frac{1}{\tau_a}\nabla_{\tilde{\bm{\eta}}}\cdot\left[\tilde{\bm{\eta}} P(\tilde{\bm{r}},\tilde{\bm{\eta}},\tilde{t}) \right]+\frac{1}{2\tau_a^2}\nabla^2_{\tilde{\bm{\eta}}} P(\tilde{\bm{r}},\tilde{\bm{\eta}},\tilde{t}) \qquad\text{and}\\
\label{eq:fpe3D_c}
\frac{\partial}{\partial \tilde{t}} \tilde{\mathcal{C}}(\tilde{\bm{r}},\tilde{t}) &= \tilde{D}_{ch}\nabla^2_{\tilde{\bm{r}}}\tilde{\mathcal{C}}(\tilde{\bm{r}},\tilde{t})\,.
\end{align}}
Equations \eqref{eq:fpe3D} and \eqref{eq:fpe3D_c} describe the space-time evolution of the probability of finding a particle and the concentration of nutrients respectively, akin to those of \cite{fodor-prl-2016,martin-pre-2021,liebchen-acm-2018}, but including the effects of thermal regelation discussed above.  Both equations contain microscopic and macroscopic scales. The regime of interest is the long time behavior, computed by deriving the effective macroscopic dynamics as described next.

\section{Results}\label{sec:results}
\subsection{\label{subsec:multiscale}Method of multiple scales}
The macroscopic length characterizing the heat flux is
\begin{equation}
L=\frac{T_m}{|\nabla T|}\,.
\label{eq:macrolength}
\end{equation}
The particle scale $l$ is such that $l << L$, and hence their ratio defines a small parameter $\epsilon$
\begin{equation}
\epsilon = \frac{l}{L}\,.
\label{eq:epsilon}
\end{equation}
We use the microscopic length $l$ and a characteristic time $\tau$, determined {\it{a posteriori}}, and introduce the following dimensionless variables
\begin{align}
&\bm{\eta}= \sqrt{\tau_a}\tilde{\bm{\eta}}\text{, } \bm{r} = \frac{\tilde{\bm{r}}}{l} \text{, } t =\frac{\tilde{t}}{\tau} \text{, }v=\frac{\tilde{v}}{u} \text{, } v_a = \frac{\tilde{v}_a}{v_{ac}}\text{, }D = \frac{\tilde{D}}{D_c}\text{, } D_{ch}=\frac{\tilde{D}_{ch}}{D_n}\text{~and~}\mathcal{C}=\frac{\tilde{\mathcal{C}}}{c_h} \,,
\label{eq:dimensionless_quanti}
\end{align}
where $\tilde{v}_a = \sqrt{\frac{2\tilde{D}_a}{\tau_a}}$ \cite{dabelow-jsm-2021,caprini-sm-2022}, $v_{ac}$ is the characteristic active velocity, $u$ and $D_c$ are the characteristic values of the regelation velocity and the premelting enhanced diffusivity respectively, and $D_n$ and $c_h$ are the characteristic values of the diffusivity and nutrient concentration respectively.  With these scalings, Eqs. \eqref{eq:fpe3D} and \eqref{eq:fpe3D_c}, become
\begin{align}
\label{eq:fpe_dimless}
P_l\frac{\partial}{\partial t} P &= -\beta_D\frac{c_h}{D_c}\nabla_{\bm{r}}\cdot\left[P\nabla_{\bm{r}} \mathcal{C} \right] -P_{e}\frac{\partial}{\partial r_3}\left[ vP\right]-P_{a}v_a\bm{\eta}\cdot\nabla_{\bm{r}}P 
&+\nabla^2_{\bm{r}}\left[DP \right]+P_{A}\nabla_{\bm{\eta}}\cdot\left[\bm{\eta}P \right]+\frac{1}{2}P_{A}\nabla^2_{\bm{\eta}}P\\ \text{and}\\
\label{eq:fpe_dimless_c}
P_{ch}\frac{\partial}{\partial t} \mathcal{C} &= D_{ch}\nabla^2_r \mathcal{C}\,, 
\end{align}
in which we have the following dimensionless numbers, 
\begin{equation}
P_{e}=\frac{ul}{D_c} \text{, } P_{a}=\frac{v_{ac}l}{D_c} \text{, } P_l=\frac{l^2}{D_c\tau} \text{, } P_{A}=\frac{l^2}{D_c\tau_a} \text{ and } P_{ch}=\frac{l^2}{D_n \tau}\,.
\end{equation}
We identify four characteristic time scales: $t_l^{\text{diff}}=l^2/D_c$, $t_l^{\text{adv}}=l/u$, $t_L^{\text{diff}}=L^2/D_c$ and $t_L^{\text{adv}}=L/u$,
associated with “microscopic” diffusion and advection on the
particle scale, $l$, and “macroscopic” diffusion and advection over the thermal length scale, $L$.
Nutrient and premelting enhanced diffusivity are taken to operate on the same time scale; $t^{\text{diff}_n}_{l,L} \sim t^{\text{diff}}_{l,L}$.
The P\' eclet number represents the ratio of the characteristic time for diffusion to that of advection, and those associated with regelation and activity are $P_{e}$ and $P_{a}$ respectively, and can be defined over both length scales, 
\begin{equation}
P_{e} = \frac{t_l^{\text{diff}}}{t_l^{\text{adv}}} \qquad \text{ and }\qquad P_{e}^L = \frac{t_L^{\text{diff}}}{t_L^{\text{adv}}}\,.
\end{equation}
The temperature gradient across the entire system drives thermal regelation and hence advection dominates on the macroscopic scale, so that $P_e^L=\mathcal{O}(1/\epsilon)$, or equivalently, $t_L^{\text{adv}}=\epsilon t_L^{\text{diff}}\sim \epsilon t_L^{\text{diff}_n}$. Whence, $P_{e}=\mathcal{O}(1)$, or equivalently, $t_l^{\text{adv}}= t_l^{\text{diff}}\sim t_l^{\text{diff}_n}$. On the macroscopic scale $P_e^L$ becomes large, as $\epsilon$ tends to zero, and thus we use the macroscopic advection time $\tau=t_L^{\text{adv}}$ as our characteristic time, so that P\' eclet numbers based on $L$ are $\mathcal{O}(1/\epsilon)$  and those based on $l$ are 
$\mathcal{O}(\epsilon)$.  In consequence, Eqs. \eqref{eq:fpe_dimless} and \eqref{eq:fpe_dimless_c} become
leading to $P_L=\mathcal{O}(1/\epsilon)$ and $P_l=\mathcal{O}(\epsilon)$, as well as $P^{ch}_L=\mathcal{O}(1/\epsilon)$ and $P^{ch}_l=\mathcal{O}(\epsilon) $. The system of Fokker-Planck equations, Eqs. \eqref{eq:fpe_dimless}-\eqref{eq:fpe_dimless_c}, becomes 
\begin{align}
\label{eq:fpe_dimless_scenaadv}
\epsilon\frac{\partial}{\partial t} P &= -\beta_D\frac{c_h}{D_c}\nabla_{\bm{r}}\cdot\left[P\nabla_{\bm{r}} c \right] -\frac{\partial}{\partial r_3}\left[ vP\right]-P_{a}v_a\bm{\eta}\cdot\nabla_{\bm{r}}P \nonumber \\
&+\nabla^2_{\bm{r}}\left[DP \right]+P_{A}\nabla_{\bm{\eta}}\cdot\left[\bm{\eta}P \right]+\frac{1}{2}P_{A}\nabla^2_{\bm{\eta}}P \qquad \text{and}\\
\label{eq:fpe_dimless_scenaadv_c}
\epsilon\frac{\partial}{\partial t} \mathcal{C} &= D_{ch}\nabla^2_r \mathcal{C}\,. 
\end{align}
Now, we let $\bm{R}=\tilde{\bm{r}}/L$ describe the macroscopic length scale, and $T=\tilde{t}/t^{\text{adv}}_l$  describe the microscopic time scale, leading to the 
following stretching of the microscopic scales;
\begin{equation}
\bm{r}=\frac{1}{\epsilon}\bm{R} \qquad \text{ and } \qquad T = \frac{1}{\epsilon}t\,.
\label{eq:sep_scale}
\end{equation}
\begin{figure}[t]
    \centering
    \includegraphics[scale=.5]{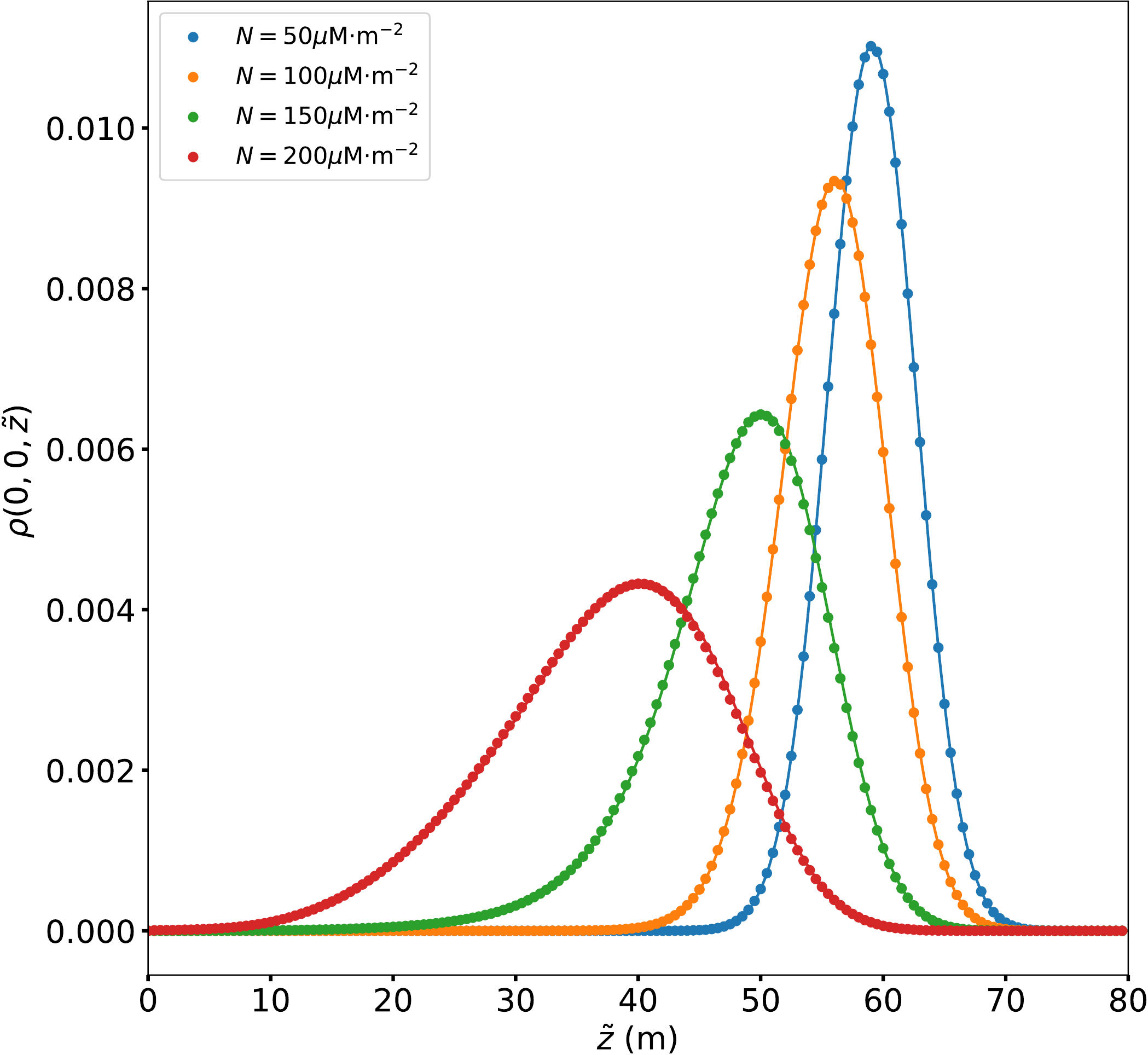}
    \caption{Consequences of bio-enhanced premelting in the absence of nutrients ($\beta_D=0$). The evolution of the probability density along the $\tilde{z}$-axis, computed from Eq. \eqref{eq:fpe_new} with $N_i=50\mu$M m$^{-2}$ and four biological enhancement factors $n\in\{1, 2, 3, 4 \}$, where $N=n N_i$.  The probability density is shown at $\tilde{x}=\tilde{y}=0$ and at time $\tilde{t}=300$ years. The analytic solution (solid lines), Eq. \eqref{eq:solu_green_funct} (see Appendix \ref{sec:appendix_large_peclet}), is compared with the numerical solution (dots) of  Eq. \eqref{eq:fpe_new}.  The particle radius is $R = 9.0 \times 10^{-6}$m.}
    \label{fig2:different_concentration}
\end{figure}
Next, we use a power series ansatz for the state variables, 
\begin{align}
    P &= P^0 +\epsilon P^1 + \epsilon^2 P^2 + \text{h.o.t.} \qquad\text{and}\\
    \mathcal{C} &= \mathcal{C}^0 +\epsilon\mathcal{C}^1 + \epsilon^2\mathcal{C} + \text{h.o.t.}\,,
\end{align}
to derive a system of equations at each order in $\epsilon$ \cite{bender-book-2013}, which for the concentration of nutrients, Eq. \eqref{eq:fpe_dimless_scenaadv_c}, are
\begin{align}
\mathcal{O}(\epsilon^0): D_{ch}\nabla_{\bm{r}}^2\mathcal{C}^0 &= 0\,,\label{eq:hiere_1c}\\
\mathcal{O}(\epsilon^1):D_{ch}\nabla_{\bm{r}}^2\mathcal{C}^1 &= \frac{\partial}{\partial T}\mathcal{C}^0-2D_{ch}\nabla_{\bm{r}}\cdot\nabla_{\bm{R}} \mathcal{C}^0\,\label{eq:hiere_2c} \qquad \text{and}\\
\mathcal{O}(\epsilon^2):D_{ch}\nabla_{\bm{r}}^2\mathcal{C}^2 &= \frac{\partial}{\partial T}\mathcal{C}^1+\frac{\partial}{\partial t}\mathcal{C}^0-2D_{ch}\nabla_{\bm{r}}\cdot\nabla_{\bm{R}}\mathcal{C}^1-D_{ch}\nabla_{\bm{R}}^2\mathcal{C}^0 \label{eq:hiere_3c}\,, 
\end{align}
shown to second order.  
We take the approach described in \cite{pavliotis-book-2008,aurell-epl-2016} to solve Eqs. \eqref{eq:hiere_1c}-\eqref{eq:hiere_3c}. 
We integrate by parts over the microscale variables $\bm{r}$ and use the periodic boundary conditions to obtain the so-called weak formulation  \cite{Chipot2009} of the leading order Eq. \eqref{eq:hiere_1c}, the solution of which relies on the following product ansatz
\begin{equation}
\mathcal{C}^0(\bm{r},\bm{R},T,t)=\zeta(\bm{r})c^0(\bm{R},T,t)\,.
\label{eq:1ansatz_c}
\end{equation} 
The existence and uniqueness of $\mathcal{C}^0$ is ensured using the Lax-Milgram theorem \cite{Chipot2009}, also known as the solvability condition or the Fredholm alternative   \cite{pavliotis-book-2008}. 
Thus, $C^0$ is constant over $\mathcal{C}^0(\bm{r},\bm{R},T,t) = \mathcal{C}^0(\bm{R},T,t)$. The solvability condition for the equation at $\mathcal{O}(\epsilon)$ is
\begin{align}
\int d\bm{r} &\left( \zeta\frac{\partial}{\partial T}c^0\right) = 0\,,
\label{eq:solvability_seondc}
\end{align}  
from which we find that $c^0$ is stationary over $T$, leading to $\mathcal{C}^0(\bm{R},T,t)=\mathcal{C}^0(\bm{R},t)$ and $\mathcal{C}^1(\bm{R},T,t)=\mathcal{C}^1(\bm{R},t)$.
Substituting $\mathcal{C}^1$ into  the $\mathcal{O}(\epsilon^2)$ Eq. \eqref{eq:hiere_3c} and using the solvability condition, gives nutrient diffusion on the macroscale as
\begin{equation}
\frac{\partial}{\partial t}c^0 = D_{ch}\nabla_{\bm{R}}^2 c^0\,,
\label{eq:fpe_dimensionlessc}
\end{equation}
showing that, as expected, the homogenization procedure is consistent with the well-known self-similar behavior of diffusion \cite{Grisha}. 
The order by order equations for the probability density function described by Eq. \eqref{eq:fpe_dimless_scenaadv} are simplified by the observation that $C^0$ and $C^1$ do not depend on $\bm{r}$, and $C^0$ only contributes at order  $\mathcal{O}(\epsilon^2)$, and hence we obtain
\begin{align}
\mathcal{O}(\epsilon^0): \mathcal{L}P^0 &= 0\,,\label{eq:hiere_1}\\
\mathcal{O}(\epsilon^1):\mathcal{L}P^1 &= \frac{\partial}{\partial T}P^0 + {\color{black}\beta_D \frac{c_h}{D_c}\nabla_{\bm{r}}\cdot \left[ P^0 \nabla_{\bm{R}} C^0\right]} + \frac{\partial}{\partial R_3}\left[ vP^0\right]+P_{a}v_a\bm{\eta}\cdot\nabla_{\bm{R}} P^0\nonumber\\
&-2\nabla_{\bm{r}}\cdot\nabla_{\bm{R}}\left[DP^0\right]\,,\label{eq:hiere_2} \qquad \text{and}\\
\mathcal{O}(\epsilon^2):\mathcal{L}P^2 &= \frac{\partial}{\partial T}P^1+\frac{\partial}{\partial t}P^0 + \beta_D \frac{c_h}{D_c}\left\{{\color{black}\nabla_{\bm{r}}\cdot \left[ P^1 \nabla_{\bm{R}} C^1\right] + \nabla_{\bm{r}}\cdot \left[ P^0 \nabla_{\bm{R}} C^0\right]} + \nabla_{\bm{R}}{\color{black}\cdot}\left[P^0\nabla_{\bm{R}} \mathcal{C}^0 \right]\right\}  \nonumber\\
&+\frac{\partial}{\partial R_3}\left[ vP^1\right]  + P_{a}v_a\bm{\eta}\cdot\nabla_{\bm{R}}P^1-2\nabla_{\bm{r}}\cdot\nabla_{\bm{R}}\left[ DP^1\right]-\nabla_{\bm{R}}^2\left[D P^0\right]\label{eq:hiere_3}\,,
\end{align}
where $\mathcal{L} = \mathcal{M}+\mathcal{Q}$, with $\mathcal{M} = -\frac{\partial}{\partial r_3}v-P_{a}v_a\bm{\eta}\cdot\nabla_{\bm{r}}+\nabla^2_{\bm{r}}D$, and $\mathcal{Q} = P_{A}\nabla_{\bm{\eta}}\cdot\bm{\eta}+\frac{P_{A}}{2}\nabla^2_{\bm{\eta}}$. 
\begin{figure}[t]
         \centering
         \includegraphics[scale=.5]{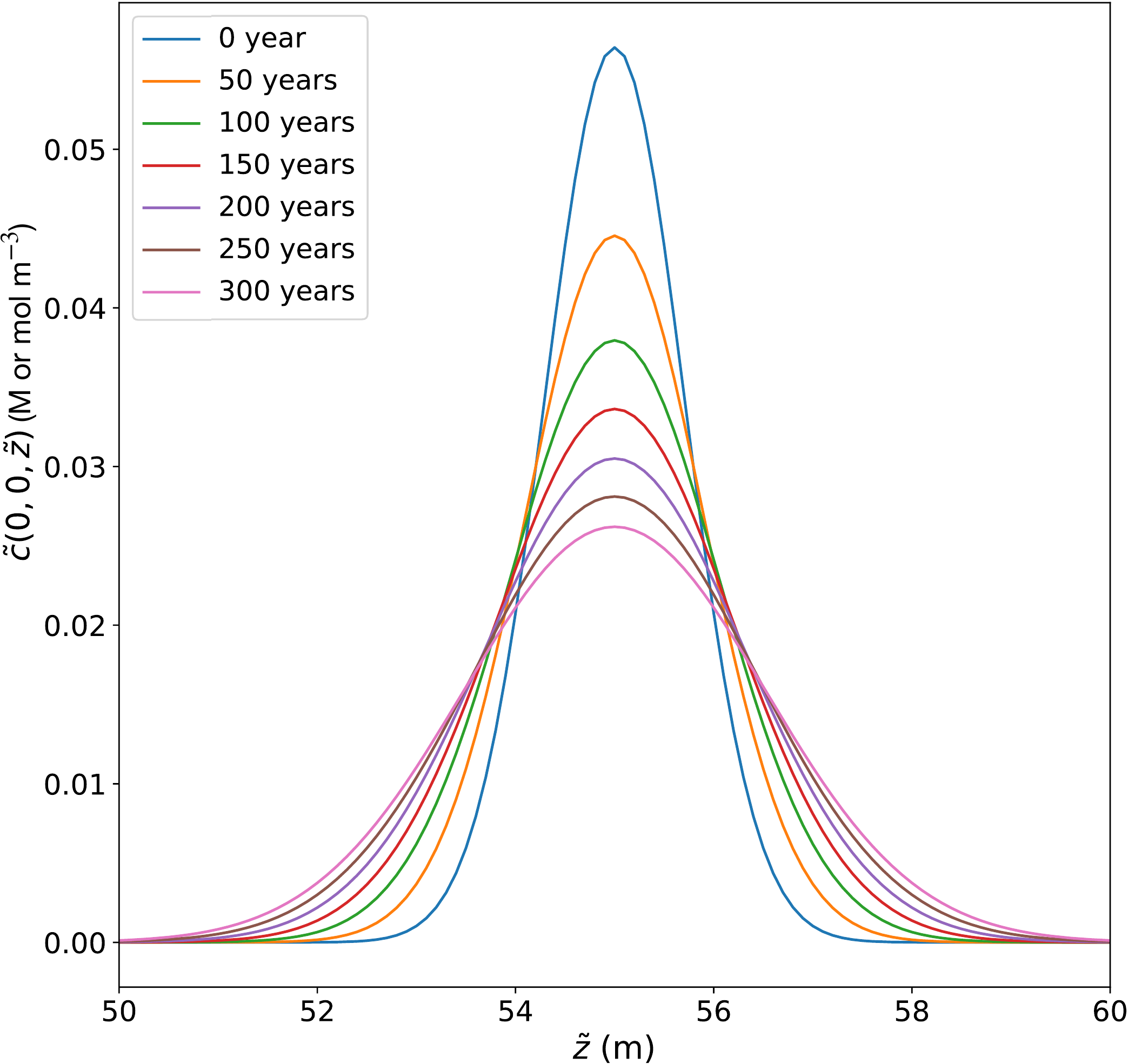}
        \caption{Evolution of the nutrient concentration (in units of M, or mol m$^{-3}$) along the $\tilde{z}$-axis, computed from Eq. \eqref{eq:fpe_new2}, at $\tilde{x}=\tilde{y}=0$. At $\tilde{t}=0$, the nutrient concentration is centered at $\tilde{z}_0=55$m and the nutrient diffusivity is $\tilde{D}_{ch}=10^{-10}$m$^2$s$^{-1}$.}
        \label{fig3:nutrient_concentration}
\end{figure}
 Finally, as shown in Appendix \ref{sec:appendix_multiple_scales_analysis}, upon substitution of $P^1$ into Eq. \eqref{eq:hiere_3} and using the solvability condition, we obtain the effective macroscopic dynamics as
\begin{align}
\label{eq:fpe_new}
    \frac{\partial}{\partial\tilde{t}}\rho &=- \beta_D\nabla_{\tilde{\bm{r}}}\cdot\left[\rho\nabla_{\tilde{\bm{r}}}\tilde{c} \right] -\frac{\partial}{\partial\tilde{z}}\left[\tilde{v}\rho \right] + \nabla^2_{\tilde{\bm{r}}}\left[(\tilde{D}_a+\tilde{D})\rho \right] \qquad \text{and}\\
    \label{eq:fpe_new2}
    \frac{\partial}{\partial\tilde{t}} \tilde{c} &= \tilde{D}_{ch}\nabla^2_{\tilde{\bm{r}}}\tilde{c}\,, 
\end{align}
which are the dimensional forms of  Eqs. \eqref{eq:hiere_3} and \eqref{eq:fpe_dimensionlessc} respectively.  These capture the long time behavior wherein the active force is treated through the effective diffusivity, which is enhanced by thermal regelation, consistent with our previous work \cite{vachier-pre-2022} and that in active matter systems generally \cite{maggi-prl-2014,bechinger-rmp-2016,caprini-sm-2018}.

Eqs. \eqref{eq:fpe_new}-\eqref{eq:fpe_new2} can be  mapped onto the well-known Keller–Segel equations for chemotaxis \cite{saha-pre-2014, pohl-prl-2014,liebchen-acr-2018,liebchen-book-2020}, where
$\rho$ is the cell density and the sign of $\beta_D$ determines whether a cell is attracted or repelled by the nutrient. Finally, when nutrients are neglected,  $\beta_D=0$, we recover our previous results \cite{marath-sm-2020,vachier-pre-2022}.\newline
\begin{figure}[t]
         \centering
         \includegraphics[scale=.58]{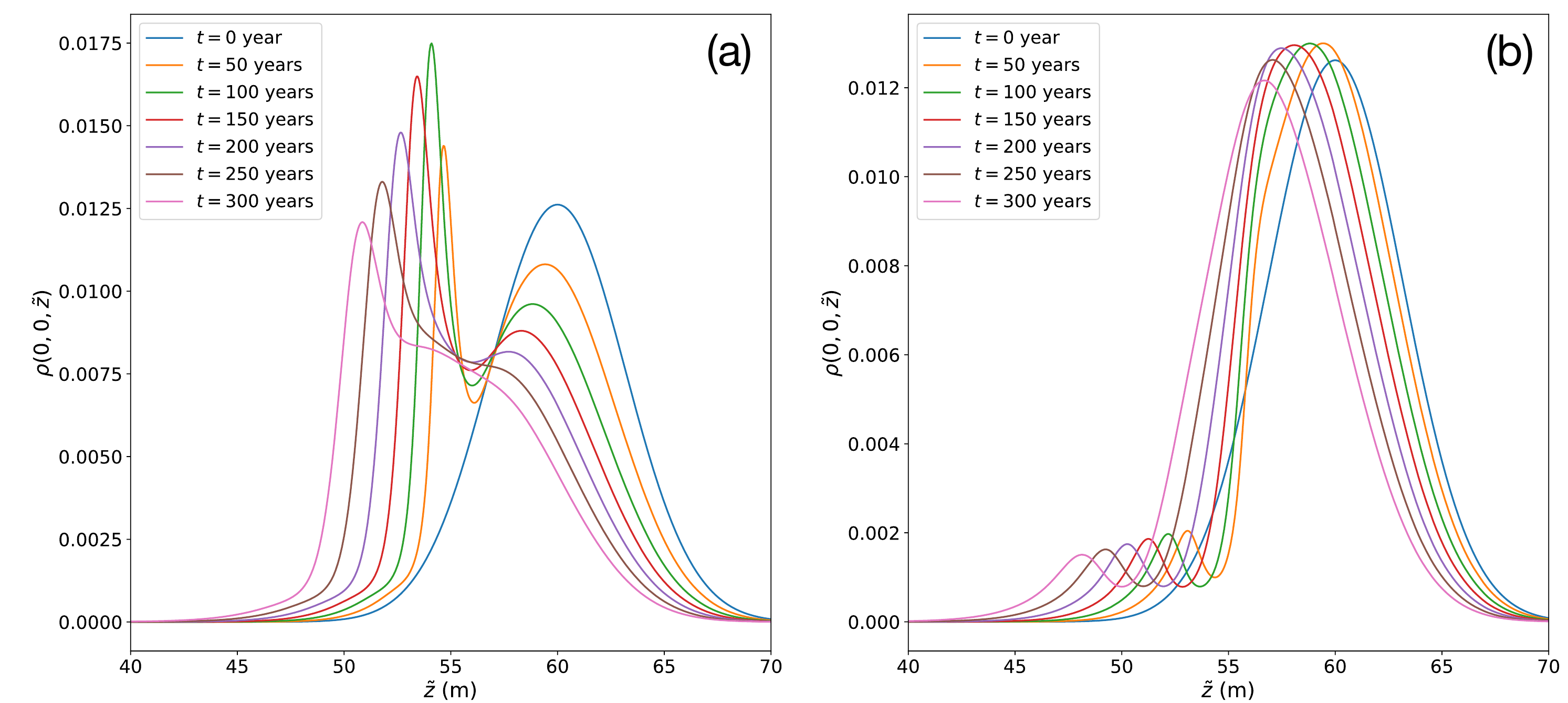}
        \caption{Effect of nutrients on the particle dynamics. Evolution of the  probability density function along the $\tilde{z}$-axis, at $\tilde{x}=\tilde{y}=0$, for two values of the chemotaxis strength $\beta_D$ {\bf{(a)}} $\beta_D=10^{-10}$ m$^{2}$M$^{-1}$s$^{-1}$ $>0$ (attractive) and {\bf{(b)}} $\beta_D=-10^{-10}$ m$^{2}$M$^{-1}$s$^{-1}$ $<0$ (repulsive).  At $\tilde{t}=0$ the distribution is centered at $\tilde{z}_0=60$m. The solution of Eq. \eqref{eq:fpe_new} is computed using a finite difference method. The particle radius is $R=10^{-6}$m, the concentration of impurities is $N=100\mu$M m$^{-2}$, the temperature gradient is $|\nabla T|=0.1$K m$^{-1}$ and the active diffusivity is $\tilde{D}_a=100\tilde{D}$.}
        \label{fig4:pdf_combin_time}
\end{figure}
Although Eq. \eqref{eq:fpe_new} has an analytical solution in the large P\' eclet number limit, which previously allowed us to study the effect of the activity (\cite[e.g., see][]{marath-sm-2020,vachier-pre-2022} or Appendix \ref{sec:appendix_large_peclet}), here we fix the activity and focus on the competition between thermal regelation and bio-locomotion that require solving Eqs. \eqref{eq:fpe_new} and \eqref{eq:fpe_new2} numerically.  We show dimensional results because of our specific interest in these processes in ice.  
\begin{figure}[t]
         \centering
         \includegraphics[scale=.8]{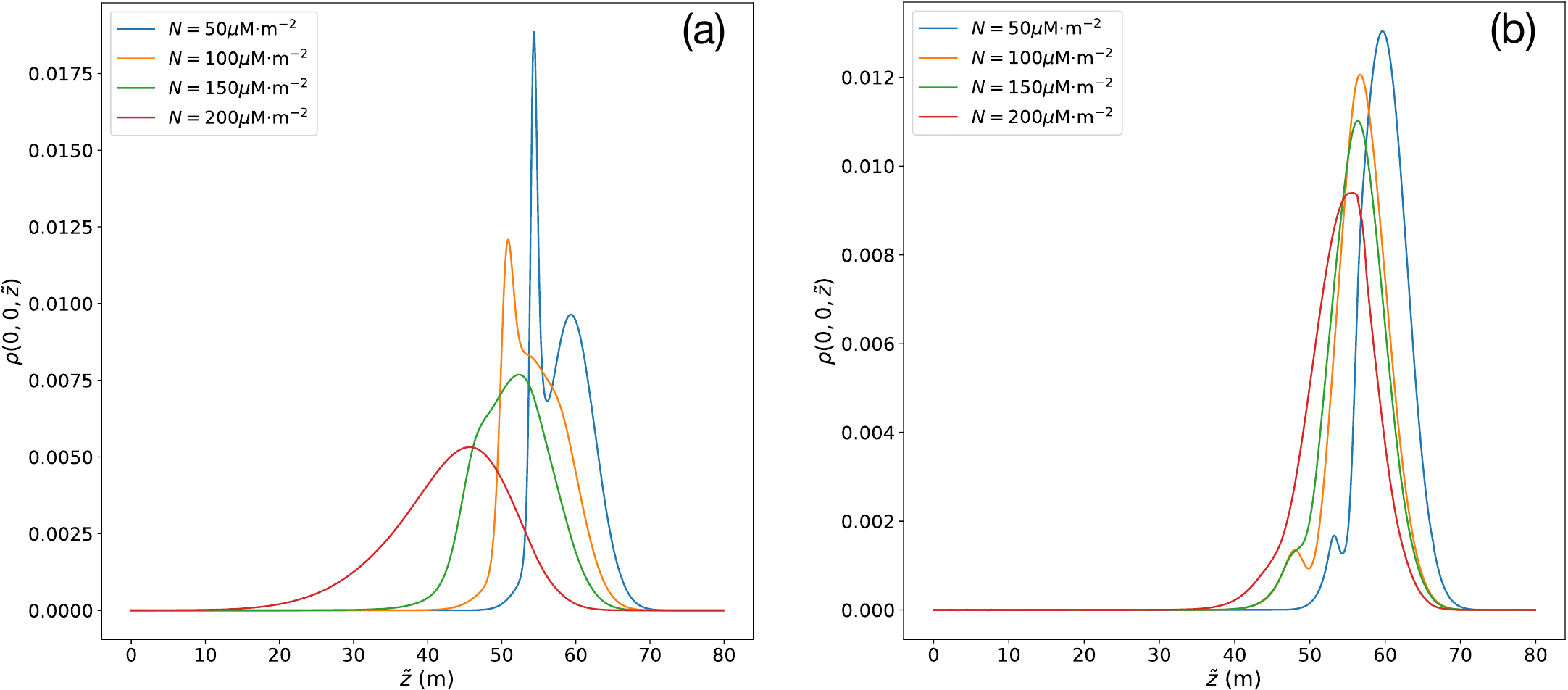}
        \caption{The combined effects of the surface concentration of impurities and nutrients. The probability density function along the $\tilde{z}$-axis, at $\tilde{x}=\tilde{y}=0$ and at $\tilde{t}=300$years, for 
         {\bf{(a)}} attractive ($\beta_D=10^{-10}$ m$^{2}$M$^{-1}$s$^{-1}$) and {\bf{(b)}} repulsive ($\beta_D=-10^{-10}$ m$^{2}$M$^{-1}$s$^{-1}$) chemotaxis, for different surface concentration of impurities $N$, computed from Eqs. \eqref{eq:fpe_new}-\eqref{eq:fpe_new2} using a finite difference method. The particle radius is $R=10^{-6}$m, the temperature gradient is $|\nabla T|=0.1$K m$^{-1}$ and the active diffusivity is $\tilde{D}_a=100\tilde{D}$.  The nutrient source is centered at $\tilde{z}$ = 55 m. }
        \label{fig5:pdf_combin_concen}
\end{figure}

\indent  In the absence of nutrients, $\beta_D=0$, Figure \ref{fig2:different_concentration} shows how the distribution of bio-particles parallel to the temperature gradient (the $\tilde{z}$-axis) is influenced by EPS/AFP production, which is modeled as a surface colligative effect.   Namely, with $N_i=50\mu$M m$^{-2}$ and four biological enhancement factors $n\in\{1, 2, 3, 4 \}$.   The active diffusivity is $\tilde{D}_a = 100 \tilde{D}$ and the particle radius is $R = 9.0 \times 10^{-6}$m. 

Figure \ref{fig3:nutrient_concentration} shows the evolution of the nutrient concentration along the $\tilde{z}$-axis computed from Eq. \eqref{eq:fpe_new2}, at $\tilde{x}=\tilde{y}=0$.  The nutrients are centered at $\tilde{z}_0=55$m at $\tilde{t}=0$ and we use a nutrient diffusivity  of $\tilde{D}_{ch}=10^{-10}$m$^2 $s$^{-1}$ \cite{theurkauff-prl-2012,jin-pnas-2017,salek-nature-2019,hokmabad-arxiv-2022}.
\newline
\indent \textcolor{black}{In order to study the effect of nutrients on bio-locomotion, we fix the interfacial concentration of impurities and vary the chemotaxis strength $\beta_D$, where nutrients either attract ($\beta_D>0$) or repel ($\beta_D<0$) the bio-particles.  Because we are interested in the situation wherein the effects of chemotaxis compete with regelation, this constrains the magnitude of $\beta_D$ as follows. We ask for what order of magnitude of $\beta_D$ are the typical chemotactic speeds approximately the same as the regelation velocity in Eq. \eqref{eq:lange_posi}.  Figure \ref{fig3:nutrient_concentration} shows the Gaussian solution of the concentration field, with a flux that becomes arbitrarily small at long times, dominated by the algebraic contribution to $ \nabla_{\tilde{\bm{r}}}\tilde{\mathcal{C}}(\tilde{\bm{r}},\tilde{t}) \propto x {t^{- 3 / 2}} \exp(-\frac{x^{2}}{t}) \sim x {t^{- 3 / 2}}$. 
For the parameters studied here, the regelation speeds are $10^{-12}-10^{-10}$m s$^{-1}$ \cite{marath-sm-2020,vachier-pre-2022}, and hence we capture this same range in $\beta_D\nabla_{\tilde{\bm{r}}}\tilde{\mathcal{C}}(\tilde{\bm{r}},\tilde{t})$, with 
$|\beta_D|=10^{-10}$ m$^{2}$M$^{-1}$s$^{-1}$, which is realized across a large time span wherein $\nabla_{\tilde{\bm{r}}}\tilde{\mathcal{C}}(\tilde{\bm{r}},\tilde{t})$ varies by several orders of magnitude.  This is also reflected in the dimensionless ratio $\beta_D\frac{c_h}{D_c}$ in Eq. \eqref{eq:fpe_dimless}.  Namely, for micron to nm scale premelted films surrounding micron sized particles  ${D_c}$ ranges from about $10^{-14}-10^{-13}$ m$^2$s$^{-1}$, and the nutrient concentration over relevant time scales has mean values ranging over $10^{-3}-10^{-2}$ M.  Therefore, $\beta_D\frac{c_h}{D_c}$ ranges from 1 to 100 and hence chemotaxis is on a similar footing to regelation under these circumstances.  For all cases considered here we use $\beta_D=\pm10^{-10}$ m$^{2}$M$^{-1}$s$^{-1}$ for attractive/repulsive chemotaxis. }
\newline
\indent \textcolor{black}{For attractive chemotaxis ($\beta_D >0$), we show in Figure \ref{fig4:pdf_combin_time} {\bf{(a)}} the dependence of $\rho(\bm{\tilde{r}},\tilde{t})$ along the $\tilde{z}$-axis parallel to the temperature gradient and at $\tilde{x}=\tilde{y}=0$, with the concentration of nutrients centered at $\tilde{z}=55$m.  For the same conditions in the absence of chemotaxis, the net displacement from low to high temperatures due to regelation is approximately 10 m \cite{vachier-pre-2022}. We see here the chemo-attractive modulation of $\rho(\bm{\tilde{r}},\tilde{t})$ during this displacement,  which ``pulls up'' the high temperature (low $\tilde{z}$) tail towards the lower temperature (large $\tilde{z}$) but higher concentration regions centered at $\tilde{z}=55$m.  The associated asymmetry depletes/attracts the low temperature regions at larger $\tilde{z}$ and concentrates the high temperature regions at smaller $\tilde{z}$, and is reflected in the evolution towards a sigmoidal region transecting the source at $\tilde{z}=55$m.  As the maximum of $\rho(\bm{\tilde{r}},\tilde{t})$ advects through the source region it first sharpens, due to the chemo-attraction from the source ``behind" it at $\tilde{z}=55$m, and then begins to spread out again because of the decay in the chemotactic gradient in time as seen in Figure~\ref{fig3:nutrient_concentration} and discussed above.}
\newline
\indent \textcolor{black}{For repulsive chemotaxis ($\beta_D < 0$), we see in Figure \ref{fig4:pdf_combin_time} {\bf{(b)}} the broad sharpening of the initial distribution in the lower temperature (large $\tilde{z}$) regions 
as it regelates/advects into the diffuse repulsive tail of nutrient field to the right of the source region centered at $\tilde{z}=55$m.  However, because the initial high temperature (small $\tilde{z}$) tail of $\rho(\bm{\tilde{r}},\tilde{t})$ interacts with the nutrient source region at $\tilde{z}=55$m, chemo-repulsion quickly drives particles towards high temperature (small $\tilde{z}$) regions, and is clearly reflected in the creation of a local maximum.  This maximum advects towards high temperature with a decaying amplitude and width due to the decay in the chemotactic gradient in time as seen in Figure~\ref{fig3:nutrient_concentration}.}
\newline
\indent \textcolor{black}{In Fig. \ref{fig5:pdf_combin_concen}, we show the combined effects of EPS/AFP surface enhancement of impurities in the absence of chemotaxis ($\beta_D=0$), as shown in Fig. \ref{fig2:different_concentration}, and the influence of chemotaxis on particle dynamics for fixed surface impurities, as shown in Fig. \ref{fig4:pdf_combin_time}.  As we vary the surface concentration of impurities we observe the same basic features as described in Figs. \ref{fig2:different_concentration} and \ref{fig4:pdf_combin_time} and hence the same physical description applies in their interpretation.  Namely, regardless of whether chemo-attraction or chemo-repulsion is operative, if the interfacial concentration of impurities $N$ is sufficiently large then the interfacial film thicknesses are sufficiently thick that thermal regelation dominates the evolution of $\rho(\bm{\tilde{r}},\tilde{t})$.   As the interfacial concentration of impurities $N$ decreases chemotaxis exerts more control on the distribution, and the basic dynamics are the same as described in Fig. \ref{fig4:pdf_combin_time}.  Because the magnitude of $\beta_D$ is fixed, and the characteristic concentration $c_h$ is $10^{-2}$ M, this $N$-dependence is simply assessed as discussed above, in terms of 
the dimensionless ratio $\beta_D\frac{c_h}{D_c}$ in Eq. \eqref{eq:fpe_dimless}.   Namely, the numerator is fixed, but as $N$ increases so too is the film thickness $d$ through Eq. \eqref{eq:thickness2}, and since ${D_c} \propto d^3$ \cite{peppin-jsp-2009}, then $\beta_D\frac{c_h}{D_c}$ decreases as $N^{-3}$, and the balance between chemotaxis control of the distribution gives way to regelation control.  The corpus of effects studied here are reflected in this basic balance and shown in Figs. \ref{fig2:different_concentration}, \ref{fig4:pdf_combin_time} and \ref{fig5:pdf_combin_concen}.}

\section{Conclusion}\label{sec:conclusion}
\indent \textcolor{black}{Micro-organisms in ice exhibit complex processes to persist and evolve in their harsh environments.  They have developed different survival strategies, such as producing exopolymeric substances or antifreeze glycoproteins, and directing their motion toward nutrients or away from waste \cite{aumack-jms-2014,price-fme-2007,stocker-mmbr-2012,bar-jrsi-2016}. We have modeled such micro-organisms using active Ornstein-Uhlenbeck particles subject to thermal regelation and biolocomotion in three dimensions. Firstly, we used a multi-scale expansion to derive the relevant coupled Fokker-Planck and diffusion equations \eqref{eq:fpe_new}-\eqref{eq:fpe_new2}. Secondly, when nutrients are neglected, and the chemotactic strength $\beta_D=0$, we model the bio-production of surface chemicals, such as exopolymeric substances or antifreeze glycoproteins, as a surface colligative effect, and find that the associated bio-enhanced thermal regelation can dominate the distribution of particles in ice.  Consistent with previous results \cite{vachier-pre-2022}, in a large P\' eclet number limit analytical solutions for the particle distributions are possible, and are consistent with the numerical solutions as shown in Fig. \ref{fig2:different_concentration}.  
Thirdly, we studied the competition between thermal regelation and biolocomotion, as function of the chemotaxis strength $\beta_D$, the interplay between which is shown in Figs. \ref{fig4:pdf_combin_time}-\ref{fig5:pdf_combin_concen}. The relative importance of chemo-attraction and chemo-repulsion to thermal regelation is captured by the dimensionless ratio $\beta_D\frac{c_h}{D_c}$.   When this ratio is large we find a complex modulation of regelation by chemotaxis, and when small, due to increased surface impurity concentration, leads to regelation dominated redistribution of particles.  We note, however, that we have not treated the process wherein nutrients themselves have a colligative effect, which would introduce a particularly complex spatio-temporal dynamics.  }

\indent \textcolor{black}{Finally, we describe settings to which our analysis is applicable.  It is of general interest to understand how particles in ice migrate in response to environmental forcing, as they are used as proxy to infer past climate \cite{miteva-book-2008,thomas-geobiol-2015,han-sr-2017}.  Moreover, bioparticles in ice migrate in response to environmental forcing, and micro-organisms play an important role in climate change \cite{mitchell-fme-2006,dutta-eee-2016,cavicchioli-nature-2019}.  For example, an increase in temperature activates algae/bacteria trapped in ice, producing chemicals that increase their mobility \cite{cavicchioli-nature-2019}.  Indeed, an increase in algae/bacteria decreases the albedo of the ice \cite{ryan-nature-2018, perini-frontiers-2019,williamson-pnas-2020}, thereby enhancing melting. 
Finally, understanding the distribution and viability of bioparticles in partially frozen media on Earth \cite[e.g.,][]{van-elementa-2018,cimoli-sr-2020} is essential in astrobiology \cite{wettlaufer-book-2010, nadeau-astrobio-2016,jones-frontiersmico-2018}.}

\section*{Conflict of Interest Statement}
The authors declare no competing interests.

\section*{Author Contributions}
J.S.W. conceived the project. J.V. implemented the theory
and performed simulations. J.S.W. and J.V. interpreted the
data and wrote the paper. All authors contributed to the discussions and the final version of the manuscript.

\section*{Funding}
This work was supported by the Swedish Research Council grant no. 638-2013-9243.  Nordita is partially supported by Nordforsk.

\section*{Acknowledgments}
We thank Matthias Geilhufe, Navaneeth Marath and Istv\' an M\' at\' a Sz\' ecs\' enyi for helpful conversations. 

\section{Appendix}
\subsection{Multiple scales Analysis}\label{sec:appendix_multiple_scales_analysis}
The solution of the leading order Eq. \eqref{eq:hiere_1} is derived by making the following product ansatz
\begin{equation}
P^0(\bm{r},\bm{R},\bm{\eta},T,t)=w(\bm{r},\bm{\eta})\rho^0(\bm{R},T,t)\,.
\label{eq:1ansatz}
\end{equation} 
Integrating by parts over the microscale variables $\bm{r}$ and $\bm{\eta}$, and using the solvability condition, the solution of Eq. \eqref{eq:hiere_1}  $P^0$ is constant over the period: $P^0(\bm{r},\bm{R},\bm{\eta},T,t)=P^0(\bm{R},\bm{\eta},T,t)$. The leading order Eq. \eqref{eq:hiere_1} becomes
\begin{equation}
\nabla_{\bm{\eta}}\cdot\left[\bm{\eta} w\right]+\frac{1}{2}\nabla^2_{\bm{\eta}}w = 0\,.
\end{equation}
Using the known result for a multi-dimensional Ornstein-Uhlenbeck process \cite{risken-book-1996}, the solution for $w$ is given by
\begin{equation}
w(\eta_1,\eta_2,\eta_3) = \prod\limits_{i=1}^3 \frac{1}{\sqrt{2\pi}}e^{\dfrac{-\eta^2_i}{2}}\,.
\label{eq:w_solution}
\end{equation}
The solvability condition for $\mathcal{O}(\epsilon)$  equation is 
\begin{align}
\int d\bm{r}d\bm{\eta} &\left( w\frac{\partial}{\partial T}\rho^0+w\frac{\partial}{\partial R_3}\left[v\rho^0\right]+wP_av_a\bm{\eta}\cdot\nabla_{\bm{R}}\rho^0\right) = 0\,,
\label{eq:solvability_seond}
\end{align}  
which depends on the leading order result, $P^0$, from which we find
\begin{equation}
\frac{\partial}{\partial T}\rho^0 = -\frac{\partial}{\partial R_3}\left[v \rho^0 \right]\,,
\label{eq:solvability_second_step1}
\end{equation}
and the $\mathcal{O}(\epsilon)$ equation becomes
\begin{equation}
\mathcal{L}P^1 = wP_av_a\bm{\eta}\cdot\nabla_{\bm{R}}\rho^0\,.
\label{eq:solvability_second_step2}
\end{equation}
We assume that
\begin{equation}
P^1=wP_av_a\bm{\alpha}\cdot\nabla_{\bm{R}}\rho^0\,,
\label{eq:2ansatz}
\end{equation}
after which we find that
\begin{equation}
\bm{\alpha} = -\frac{1}{P_{A}}\bm{\eta}\,.
\label{eq:solution_1auxiliary}
\end{equation}
Substitution of $P^1$ into the $\mathcal{O}(\epsilon^2)$ equation and using the solvability condition, we obtain
\begin{equation}
\frac{\partial}{\partial t}\rho^0 = \frac{P_a^2v_a^2}{2P_{A}}\nabla^2_{\bm{R}}\rho^0 - \beta_D\frac{c_h}{D_c}\nabla_{\bm{R}}\cdot\left[\rho^0\nabla_{\bm{R}} c^0 \right]+\nabla_{\bm{R}}^2\left[D\rho^0 \right]\,,
\label{eq:fpe_dimensionless}
\end{equation}
and in dimensional form, we have
\begin{equation}
\frac{\partial}{\partial\tilde{t}}\rho = - \beta_D\nabla_{\tilde{\bm{r}}}\cdot\left[\rho\nabla_{\tilde{\bm{r}}}\tilde{c} \right] -\frac{\partial}{\partial\tilde{z}}\left[\tilde{v}\rho \right] + \nabla^2_{\tilde{\bm{r}}}\left[(\tilde{D}_a+\tilde{D})\rho \right]\,.
\end{equation}
\subsection{Solution of the Fokker-Planck equation in the
large P\' eclet limit}\label{sec:appendix_large_peclet}
The analytic solution of Eq. \eqref{eq:fpe_new} in the large P\' eclet number limit follows by expanding the probability density function, $\rho$ perturbatively \cite{celani-pnas-2010}, as
\begin{equation}
\rho = \rho_0+\rho_1\,,
\label{eq:expansion}
\end{equation}
with $\rho_1=\mathcal{O}(\tilde{c})\rho_0$. The associated system of equations is 
\begin{align}
\frac{\partial}{\partial\tilde{t}}\rho_0 +\frac{\partial}{\partial\tilde{z}}\left[\tilde{v}\rho_0 \right] - \nabla^2_{\tilde{r}}\left[(\tilde{D}_a+\tilde{D})\rho_0 \right] &= 0\,,\\
\frac{\partial}{\partial\tilde{t}}\rho_1 +\frac{\partial}{\partial\tilde{z}}\left[\tilde{v}\rho_1 \right] - \nabla^2_{\tilde{r}}\left[(\tilde{D}_a+\tilde{D})\rho_1 \right] &= - \beta_D\nabla_{\tilde{\bm{r}}}\cdot\left[\rho_0\nabla_{\tilde{\bm{r}}}\tilde{c} \right]\,.
\end{align}
To derive the solutions for $\rho_0$ and $\rho_1$, we use the Green's function $G(\tilde{\bm{r}},\tilde{t}|\tilde{\bm{r}_0},\tilde{t}_0)$ \cite{kheifets-report-1982} which satisfies
\begin{equation}
\frac{\partial}{\partial\tilde{t}}G+\frac{\partial}{\partial\tilde{z}}\left[\tilde{v}G \right]-\nabla^2_{\tilde{\bm{r}}}\left[(\tilde{D}_a+\tilde{D})G \right] =0\,.
\label{eq:green_func}
\end{equation}
The solution of Eq.\eqref{eq:green_func} was derived in the large P\' eclet number limit in \cite{vachier-pre-2022}, and is
{\small
\begin{align}
\rho_0=G(\tilde{\bm{r}},\tilde{t}|\tilde{\bm{r}_0},\tilde{t}_0) &= \frac{\tilde{z}^3}{(\tilde{z}')^{3/4}}\exp\left(-\frac{\left[(\tilde{z}')^{1/4}-\tilde{z}_0\right]^2}{20+4\tilde{D}_a(\tilde{t}-\tilde{t}_0)} \right) \exp\left[ -\frac{((\tilde{x}-\tilde{x}_0)^2+(\tilde{y}-\tilde{y}_0)^2)}{\left( 1+4\frac{\tilde{D}(\tilde{z})}{\tilde{v}(\tilde{z})}\left[(\tilde{z}')^{1/4}-\tilde{z}\right]+4\tilde{D}_a(\tilde{t}-\tilde{t}_0)\right)} \right]\nonumber\\
&\times \frac{1}{2\sqrt{5\pi}\left(1+4\frac{\tilde{D}(\tilde{z})}{\tilde{v}(\tilde{z})}\left[(\tilde{z}')^{1/4}-\tilde{z}\right]+4\tilde{D}_a(\tilde{t}-\tilde{t}_0)  \right)}\,.
\label{eq:solu_green_funct}
\end{align}}
Given the Green's function  the formal solution of Eq.\eqref{eq:fpe_new} is 
\begin{align}
\rho(\tilde{\bm{r}},\tilde{t}) &= G(\tilde{\bm{r}},\tilde{t}|\tilde{\bm{r}_0},\tilde{t}_0) - \beta_D\int\limits_{\tilde{t}_0}^{\tilde{t}}d\tau\int\limits d\tilde{\bm{r}}_1 G(\tilde{\bm{r}},\tilde{t},|\tilde{\bm{r}}_1,\tau)\left[\nabla_{\tilde{\bm{r}}_1}\left(G(\tilde{\bm{r}}_1,\tau|\tilde{\bm{r}}_0,\tilde{t}_0)\nabla_{\tilde{\bm{r}}_1}\tilde{c}(\tilde{\bm{r}}_1,\tau|\tilde{\bm{r}}_0,\tilde{t}_0) \right) \right]\,,
\label{eq:fpe_new_solution}
\end{align}
with
\begin{equation}
    \tilde{c}(\tilde{\bm{r}},\tilde{t}|\tilde{\bm{r}}_0,\tilde{t}_0) = \frac{1}{\left(4\pi\tilde{D}_{ch}(\tilde{t}-\tilde{t}_0)\right)^{\frac{3}{2}}}\exp\left[-\frac{((\tilde{x}-\tilde{x}_0)^2+(\tilde{y}-\tilde{y}_0)^2+(\tilde{z}-\tilde{z}_0)^2)}{4\tilde{D}_{ch}(\tilde{t}-\tilde{t}_0)} \right] \,.
    \label{eq:C_app}
\end{equation}
At $\tilde{t}_0=0$ and $\tilde{x}_0=\tilde{y}_0=0$, the initial distribution is, to leading order $\rho_0$, given by
\begin{equation}
\rho_0(\tilde{\bm{r}},\tilde{t}=0) = \frac{1}{2\sqrt{5\pi}}\exp\left[-\frac{(\tilde{z}-\tilde{z}_0)^2}{20}-(\tilde{x}^2+\tilde{y}^2)\right]\,.
\label{eq:solution_fpe_t0}
\end{equation}
Equations \eqref{eq:fpe_new_solution} and \eqref{eq:C_app} give the analytic solution to Eqs. \eqref{eq:fpe_new}-\eqref{eq:fpe_new2}, with initial distribution given by Eq. \eqref{eq:solution_fpe_t0}.  When the nutrients are neglected we recover our previous result \cite{vachier-pre-2022}.

\bibliography{biblio.bib, biblioResponse}


\end{document}